\documentclass{article}
\usepackage{spconf,amsmath,graphicx,hyperref, amssymb}
\usepackage{multirow}
\usepackage{booktabs}
\usepackage{url}
\usepackage{bm} 
\usepackage[nolist, nohyperlinks]{acronym}
\usepackage{multirow}

\title{Do We Need EMA for Diffusion-Based Speech Enhancement? \\Toward a Magnitude-Preserving Network Architecture}
\name{Julius Richter \qquad Danilo de Oliveira \qquad Timo Gerkmann}
\address{University of Hamburg, Germany, Department of Informatics, Signal Processing Group}

\newcommand{\lsigma}{\overset{\leftarrow}{\phantom{.}\sigma_t}\!\phantom{}^{2}}
\newcommand{\rsigma}{\overset{\rightarrow}{\phantom{.}\sigma_t}\!\phantom{}^{2}}

\begin{document}
\ninept

\begin{acronym}
\acro{sgm}[SGM]{score-based generative model}
\acro{snr}[SNR]{signal-to-noise ratio}
\acro{STFT}[STFT]{short-time Fourier transform}
\acro{iSTFT}[iSTFT]{inverse short-time Fourier transform}
\acro{SDE}[SDE]{stochastic differential equation}
\acro{ODE}[ODE]{ordinary differential equation}
\acro{ou}[OU]{Ornstein-Uhlenbeck}
\acro{PESQ}[PESQ]{Perceptual Evaluation of Speech Quality}
\acro{tf}[T-F]{time-frequency}
\acro{SNR}{signal-to-noise ratio}
\acro{SDR}{signal-to-distortion ratio}
\acro{SI-SDR}{scale invariant signal-to-distortion ratio}
\acro{ESTOI}{Extended Short-Term Objective Intelligibility}
\acro{EMA}[EMA]{exponential moving average}
\acro{SB}[SB]{Schrö\-din\-ger bridge}
\acro{EDM2SE}[EDM2SE]{Elucidating Diffusion Models 2 for Speech Enhancement}
\acro{EDM2}[EDM2]{Elucidating Diffusion Models 2}
\acro{ADM}[ADM]{ablated diffusion model}
\acro{MP-ADM}[MP-ADM]{magnitude-preserved ablated diffusion model}
\acro{FID}[FID]{Fréchet inception distance}
\end{acronym}

\maketitle

\begin{abstract}
We study diffusion-based speech enhancement using a Schrödinger bridge formulation and extend the EDM2 framework to this setting. 
We employ time-dependent preconditioning of network inputs and outputs to stabilize training and explore two skip-connection configurations that allow the network to predict either environmental noise or clean speech. 
To control activation and weight magnitudes, we adopt a magnitude-preserving architecture and learn the contribution of the noisy input within each network block for improved conditioning. 
We further analyze the impact of exponential moving average (EMA) parameter smoothing by approximating different EMA profiles post training, finding that, unlike in image generation, short or absent EMA consistently yields better speech enhancement performance. 
Experiments on VoiceBank-DEMAND and EARS-WHAM demonstrate competitive signal-to-distortion ratios and perceptual scores, with the two skip-connection variants exhibiting complementary strengths. 
These findings provide new insights into EMA behavior, magnitude preservation, and skip-connection design for diffusion-based speech enhancement.
\end{abstract}

\begin{keywords}
Speech enhancement, diffusion models, Schrö\-dinger bridge, exponential moving average, EMA
\end{keywords}

\section{Introduction}

Diffusion-based generative models have emerged as powerful tools for speech enhancement, learning clean speech posteriors conditioned on noisy inputs~\cite{richter2023speech}.
They excel at mitigating a wide range of communication artifacts, including background noise, reverberation, bandwidth limitation, codec artifacts, and packet loss~\cite{richter2024causal}, while offering flexibility for integration with model-based approaches~\cite{lemercier2025diffusion}.

Diffusion models, or score-based generative models, operate by progressively removing noise through a learned reverse process~\cite{ho2020denoising,song2021sde}.
Various stochastic processes have been explored for speech enhancement, including the Ornstein-Uhlenbeck process~\cite{richter2023speech}, the Brownian bridge~\cite{lay23_interspeech}, and the \ac{SB}~\cite{jukic2024schr}.
In all cases, we train a neural network to separate clean speech from environmental and Gaussian noise introduced during the forward process.  
Typical architectural choices include a U-Net structure \cite{ronneberger2015u}, optionally augmented with self-attention layers \cite{vaswani2017attention}.

Recent work has highlighted the importance of preconditioning to stabilize training and prevent large gradients~\cite{karras2022elucidating,gonzalez2024investigating}, yet such techniques have not been fully explored for diffusion bridges~\cite{chen2021likelihood,liu2023image}, particularly those applied to speech~\cite{jukic2024schr,richter2025investigating}.
Similarly, uncontrolled magnitude growth in network activations and weights can degrade performance~\cite{karras2024analyzing}, motivating the use of magnitude-preserving architectures~\cite{richter2025investigating,deoliveira2024non,de2025lipdiffuser}.
Another underexplored aspect is the role of \ac{EMA} parameter smoothing. While long \ac{EMA} horizons improve mode coverage in image generation, their benefit for speech enhancement remains unclear.

To address these gaps, we extend prior work on diffusion training dynamics to speech enhancement and systematically investigate the effects of magnitude preservation, skip-connection design, and EMA behavior. 

\noindent Our main contributions are as follows:
\begin{itemize}
    \item \textbf{Extending the EDM2 framework to SE.} We build on the improved training dynamics of \ac{EDM2}~\cite{karras2024analyzing} by adopting a magnitude-preserving architecture with time-dependent input/output preconditioning for stable training of Schrödinger bridge diffusion models.
    
    \item \textbf{Skip-connection design.} We introduce and compare two skip-connection configurations enabling the neural network to predict either environmental noise or clean speech, revealing complementary advantages across SI-SDR and perceptual quality metrics.
    
    \item \textbf{EMA ablation.} We systematically approximate and evaluate different EMA profiles post-training and show, for the first time, that shorter or even absent EMA horizons outperform longer ones in diffusion-based speech enhancement.
    
    \item \textbf{Comprehensive evaluation.} We demonstrate competitive results on VoiceBank-DEMAND and EARS-WHAM and provide code, audio examples, and pretrained checkpoints for reproducibility\footnote{\url{https://github.com/sp-uhh/edm2se}}.
\end{itemize}

Our experiments on VoiceBank-DEMAND~\cite{valentini2016investigating} and EARS-WHAM~\cite{richter2024ears} show that shorter or even absent EMA lengths consistently outperform long EMA horizons, and that the two skip-connection designs trade off between higher SDR and better perceptual scores.
Together, these findings provide new insights into EMA behavior, skip-connection design, and stable training dynamics for diffusion-based speech enhancement.
By exploring the \ac{MP-ADM} architecture~\cite{karras2024analyzing} in the EDM2 framework, we provide an alternative to the commonly employed NCSN++ architecture~\cite{song2021sde} in generative SE.

\section{Background}\label{sec:related_work}

\subsection{Schrödinger Bridge for Speech Enhancement}\label{sec:sbse}

The Schrödinger bridge (SB), first introduced in quantum mechanics~\cite{schrodinger1932theorie} and later connected to optimal transport~\cite{leonard2014survey}, formulates a stochastic optimal control problem interpolating between two distributions.  
It is commonly formulated as  
\begin{equation}  
  \label{eq:sb}  
  \min_{\mathbb Q \in \Pi(p_{x}, p_{y})}  
  D_\text{KL}\!\bigl(\mathbb Q \,\|\, \mathbb P \bigr),  
\end{equation}  
where \(\mathbb Q\) and \(\mathbb P\) denote controlled and uncontrolled path measures, respectively, and \(\Pi(p_{x}, p_{y})\) is the set of path measures whose boundary marginals are given by the distributions \(p_{x}\) and \(p_{y}\). 
This yields the control formulation
\begin{equation}\label{eq:optimal_control_formulation}
\begin{gathered}
    \min_{\mathbf u} \mathbb{E}\!\left[ \int_0^1 \!\!\tfrac{1}{2} \|\mathbf u(\mathbf x_t, t)\|^2 \mathrm dt \right] \\
    \text{s.t.}\quad \mathrm d\mathbf x_t = [\mathbf f(\mathbf x_t, t) + \mathbf u(\mathbf x_t, t)] \mathrm  dt + g(t) \mathrm  d \mathbf w_t, \\ 
    \mathbf x_0\sim p_x, \quad \mathbf x_1\sim p_y
\end{gathered}
\end{equation}
where $\{\mathbf{x}_t\}$ is a stochastic process governed by the \ac{SDE} with drift  $\mathbf f$, diffusion coefficient $g$, and $\mathbf w_t$ a standard Wiener process. 
The control function $\mathbf u: \mathbb R^d \times [0,1] \to \mathbb R^d$ specifies 
the control input at each state and time.  
Applying the Cole-Hopf transform to the necessary conditions of  \eqref{eq:optimal_control_formulation} leads to a system of coupled PDEs~\cite{chen2021likelihood},
\begin{equation}
\label{eq:SB-PDE}
\begin{gathered}
\begin{cases}
    \partial_t \Psi_t =-\nabla \Psi_t^\top \mathbf f-\frac{1}{2}\operatorname{Tr}(g^2\nabla^2 \Psi_t), \\
    \partial_t \widehat{\Psi}_t =-\nabla\!\cdot\!(\widehat{\Psi}_t \mathbf f)+\frac{1}{2}\operatorname{Tr}(g^2\nabla^2\widehat{\Psi}_t),
\end{cases} \\
\text{s.t.}\quad \Psi_0\widehat\Psi_0=p_{x},\quad\Psi_1\widehat\Psi_1=p_{y},
\end{gathered}
\end{equation}
where $\Psi_t$ and $\widehat{\Psi}_t$ are called time-varying (Schrödinger) potentials that are linked through Nelson’s identity  $\Psi_t\widehat\Psi_t=p_t$~\cite{nelson1967dynamical}.  
Then, the optimal control is  $\mathbf u^*(\mathbf x_t,t)=g(t)^2\nabla\log\Psi_t(\mathbf x_t)$, yielding the forward and reverse \acp{SDE}:
\begin{align}
  \mathrm d\mathbf{x}_t &= \!\left[\mathbf f(\mathbf x_t, t)+g^2(t)\nabla_{\mathbf x_t}\log\Psi_t\right]\!\mathrm dt+g(t)\mathrm d\mathbf w_t, \label{eq:sb_forward_sde} \\
  \mathrm d\mathbf{x}_t &= \!\left[\mathbf f(\mathbf x_t, t)-g^2(t)\nabla_{\mathbf x_t}\log\widehat{\Psi}_t\right]\!\mathrm dt+g(t)\mathrm d\bar{\mathbf w}_t. \label{eq:sb_reverse_sde}
\end{align}
For Gaussian boundary marginals, closed-form solutions exist~\cite{bunne2023schrodinger}.  
Assuming a drift $\mathbf f = 0$, the SB marginal conditioned on $(\mathbf x_0, \mathbf y)$ is
\begin{equation}
\label{eq:marginal_dist}
p_t(\mathbf x_t|\mathbf x_0,\mathbf y)=\mathcal{N}\!\left(\mathbf x_t;\bm{\mu}_t(\mathbf x_0, \mathbf y),\sigma_{t}^2\mathbf I\right),
\end{equation}
with mean and variance
\begin{align}
\bm{\mu}_t (\mathbf x_0, \mathbf y) &= w_x(t)\mathbf x_0+w_y(t)\mathbf y,\\
\sigma_{t}^2 &=\frac{\rsigma \lsigma}{\rsigma + \lsigma}
\end{align}
where $\rsigma\!=\!\int_0^t g^2(\tau) \mathrm d \tau$ and $\lsigma\!=\!\int_t^1 g^2(\tau) \mathrm d \tau$ are variances accumulated from either side, and $w_x(t) = \lsigma / (\rsigma + \lsigma)$ and $w_y(t) = \rsigma / (\rsigma + \lsigma)$~\cite{liu2023image}.

Following previous work~\cite{richter2023speech, jukic2024schr, welker2022speech}, we define the diffusion process in the \ac{STFT} domain, where 
\(\mathbf{x}_0\) and \(\mathbf{y}\) denote the clean and noisy speech, respectively.  
The real and imaginary components are treated as separate channels, and the time–frequency coefficients are flattened into vectors in $\mathbb R^d$, where $d$ denotes the number of time–frequency bins.
To train an SB model, a denoiser $D_\theta$ is optimized using a data prediction loss~\cite{jukic2024schr}:
\begin{equation}
\hspace{-2px}\mathcal{J}_\text{SB}(\theta)=\mathbb E\!\left[\|D_\theta(\mathbf x_t,\mathbf y,t)-\mathbf x_0\|_2^2+\alpha\|\underline{\hat{\mathbf x}}_\theta(\mathbf x_t,\mathbf y,t)-\underline{\mathbf x}_0\|_1\right]
\end{equation}
where the time-domain signals $\underline{\hat{\mathbf x}}_\theta(\mathbf x_t,\mathbf y,t) = \operatorname{iSTFT}(D_\theta(\mathbf x_t,\mathbf y,t))$ and $\underline{\mathbf x}_0 \!=\! \operatorname{iSTFT}(\mathbf x_0)$ are obtained via the \ac{iSTFT}, and $\alpha$ weights the time-domain loss term.
At inference, the reverse SDE in~\eqref{eq:sb_reverse_sde} is solved with an ODE or SDE sampler; we use the ODE sampler following~\cite{jukic2024schr}.

\subsection{Magnitude-Preserving Learned Layers}\label{sec:magnitude_preserving_layers}

Diffusion models can suffer from uncontrolled growth in activations and weights~\cite{karras2024analyzing}.  
Magnitude-preserving layers address this by normalizing weights and activations to keep their variance constant~\cite{karras2024analyzing, salimans2016weight}.  
For a linear or convolutional layer, each channel-specific weight vector $\mathbf v$ is normalized:
\begin{equation}\label{eq:weight_normalization}
\tilde{\mathbf v}=\frac{\mathbf v}{\|\mathbf v\|_2+\epsilon},
\end{equation}
with a small constant $\epsilon$ added for numerical stability to avoid division by zero.
After each gradient step, the weight vector is rescaled such that its norm satisfies $\|\tilde{\mathbf v}\|_2 = \sqrt{k}$, where $k$ is the dimensionality of $\tilde{\mathbf{v}}$.  
This preserves gradient flow and stabilizes training~\cite{karras2024analyzing}.

\subsection{Post-training Exponential Moving Average (EMA)}

\ac{EMA} maintains a running average of the parameters over the training iterations according to
\begin{equation}
\hat{\theta}_\beta^{(i)}=\beta\hat{\theta}_\beta^{(i-1)}+(1-\beta)\theta^{(i)}\,,
\end{equation}
where $i$ denotes the current training step and \(\beta\) specifies the momentum of the running average, typically set close to 1.
Karras et al.~\cite{karras2024analyzing} propose a power-law \ac{EMA} with step-dependent $\beta_\gamma^{(i)}$ defined as
\begin{equation}
\hat{\theta}_\gamma^{(i)}=\beta_\gamma^{(i)}\hat{\theta}_\gamma^{(i-1)}+(1-\beta_\gamma^{(i)})\theta^{(i)},
\quad\beta_\gamma^{(i)}=(1-1/i)^{\gamma+1}\,,
\end{equation}
where the constant $\gamma$ controls the sharpness of the profile.
The \ac{EMA} profile is characterized by its relative standard deviation $\sigma_\text{rel}$, representing the peak width relative to training
duration.

They also introduced a post-training approximation method to reconstruct arbitrary EMA profiles from stored snapshots, enabling dense evaluation of EMA length.  
While longer EMA horizons are known to improve diversity in image generation, their impact on speech enhancement remains largely unexplored.

\section{Method}\label{sec:method}

We call our diffusion-based speech enhancement
framework \ac{EDM2SE}, which is based on the official implementation\footnote{\url{https://github.com/NVlabs/edm2}} of \ac{EDM2}~\cite{karras2024analyzing}.
We formulate diffusion-based speech enhancement as an SB process (Sec.~\ref{sec:sbse}) in the STFT domain with amplitude compression as in~\cite{richter2023speech}, treating the real and imaginary parts as separate channels and flattening time-frequency coefficients into vectors in \(\mathbb{R}^d\). 

Let $\mathbf{x}_0$ denote the clean speech signal and $\mathbf{y}$ the corresponding noisy speech mixture, modeled as 
\begin{equation}
\label{eq:mixture_model}
 \mathbf{y} = \mathbf{x}_0 + \mathbf{n},
\end{equation}
where $\mathbf{n}$ represents the additive environmental noise.  
We assume that the clean speech \(\mathbf{x}_0\) and the environmental noise \(\mathbf{n}\) are statistically independent, with variances 
\(\operatorname{Var}(\mathbf{x}_0)=\sigma_{x}^2\) and \(\operatorname{Var}(\mathbf{n})=\sigma_{n}^2\), which we compute from the training data.
Moreover, we consider the \ac{SB} conditional marginal distribution in~\eqref{eq:marginal_dist} with mean \(\bm{\mu}_t(\mathbf{x},\mathbf{y})\) and variance \(\sigma_{_t}^2\).

\subsection{Preconditioning}

Preconditioning rescales the network inputs and outputs at each process time \(t\) to stabilize optimization~\cite{karras2022elucidating}. Given
\begin{equation}
\mathbf{x}_t = \bm{\mu}_t(\mathbf{x}_0,\mathbf{y}) + \sigma_t\mathbf{z}, \qquad \mathbf{z}\sim\mathcal{N}(\mathbf{0},\mathbf{I}),
\end{equation}
we train a denoiser \(D_\theta\) with the time-weighted objective
\begin{equation}
\label{eq:loss}
\mathcal{J}(\theta) = \mathbb{E}_{t,\mathbf{x}_0,\mathbf{y},\mathbf{z}}\!\left[\lambda(t)\bigl\|D_\theta(\mathbf{x}_t, \mathbf y,t)-\mathbf{x}_0\bigr\|_2^2\right],
\end{equation}
where $t\sim\mathcal{U}(t_{\text{eps}},1)$ and $\lambda(t)$ is a time-dependent weight.
We parameterize $D_\theta$ as
\begin{equation}
\label{eq:preconditioning}
D_\theta(\mathbf{x}_t, \mathbf{y}, t)
= c_{\text{s}}\mathbf{x}_t + c_{\text{out}}(t)
F_\theta(c_{\text{in}}(t)\mathbf{x}_t, c_{\text{in}}(1)\mathbf{y}, t),
\end{equation}
where \(F_\theta\) is the neural network, \(c_{\text{in}}\) and \(c_{\text{out}}\) are time-dependent input/output scalings, and \(c_{\text{s}}\) is a constant defining the skip connection applied to \(\mathbf{x}_t\).
The conditioner uses \(c_{\text{in}}(1)\), corresponding to the normalization that yields unit variance for \(\mathbf{x}_t\) at \(t=1\). 

Substituting \eqref{eq:preconditioning} into \eqref{eq:loss} yields
\begin{equation}
\mathcal{J}(\theta)
= \mathbb{E}\Bigl[\, \underbrace{\lambda(t)\,c_{\text{out}}(t)^2}_{w(t)}\,
\bigl\|F_\theta(\cdot)-F_{\text{target}}(\cdot)\bigr\|_2^2 \Bigr], 
\end{equation}
\vspace{-1.0em}
\begin{equation}
F_{\text{target}}(\mathbf{x}_0,\mathbf{x}_t,t)
= (\mathbf{x}_0-c_{\text{s}}\,\mathbf{x}_t) / c_{\text{out}}(t),
\end{equation}
where \(w(t)\) is the effective weight and \(F_\theta\) learns a normalized target. \\

\noindent\textbf{Input Scaling:}\;  We require unit-variance inputs to \(F_\theta\), such that
\begin{align}
\operatorname{Var}\!\left(c_{\text{in}}(t)\mathbf{x}_t\right) \overset{!}{=} 1
\;\;\Leftrightarrow\;\;
c_{\text{in}}(t)^2\,\operatorname{Var}\!\left(\bm{\mu}_t+\sigma(t)\mathbf{z}\right) = 1.
\end{align}
Using \eqref{eq:mixture_model}, we have 
\(\bm{\mu}_t = (w_{x}(t)+w_{y}(t))\mathbf{x}_0 + w_{y}(t)\mathbf{n}\); hence
\begin{equation}
\label{eq:input_scaling}
c_{\text{in}}(t) = 1 / \sqrt{\,\bigl(w_{x}(t)+w_{y}(t)\bigr)^2 \sigma_{x}^2
+ w_{y}(t)^2 \sigma_{n}^2
+ \sigma_t^2}\,.
\end{equation}

\noindent\textbf{Output Scaling:}\; We require  $\operatorname{Var}\!\left(F_{\text{target}}(\cdot, \cdot, t)\right) \overset{!}{=} 1$. Thus, we get
\begin{align}
\label{eq:output_scaling}
c_{\text{out}}(t)^2
&= (1 - c_{\text{s}}(w_{x}(t)+w_{y}(t)))^2 \sigma_{x}^2 + c_{\text{s}}^2 w_{y}(t)^2 \sigma_{n}^2
+ c_{\text{s}}^2 \sigma_t^2.
\end{align}

\noindent\textbf{Skip Scaling:}\; We explore \(c_{\text{s}}\!\in\!\{0,1\}\).
For \(c_{\text{s}}\!=\!1\) (noise prediction),
\begin{equation}
\label{eq:output_scaling_1}
c_{\text{out},n}(t)
= \sqrt{\,\bigl(1 - w_{x}(t) - w_{y}(t)\bigr)^2 \sigma_{x}^2
+ w_{y}(t)^2 \sigma_{n}^2
+ \sigma_t^2\,}.
\end{equation}
For \(c_{\text{s}}=0\) (clean speech prediction), we get
\begin{equation}
\label{eq:output_scaling_0}
c_{\text{out},x}(t) \;=\; \sigma_{x}.
\end{equation}

\noindent\textbf{Loss Weighting:}\; We set \(w(t) = 1\), i.e., \(\lambda(t)=1/c_{\text{out}}(t)^2\).
For noise prediction and clean speech prediction, respectively, we get
\begin{align}
\lambda_n(t) &= 1 / ( (1 - w_{x}(t) - w_{y}(t))^2 \sigma_{x}^2
+ w_{y}(t)^2 \sigma_{n}^2 + \sigma_t^2), \\
\lambda_x(t) &= \sigma_{x}^{-2}.
\end{align}

\subsection{Network Architecture}

\begin{figure}
    \centering
    \includegraphics[width=0.80\linewidth]{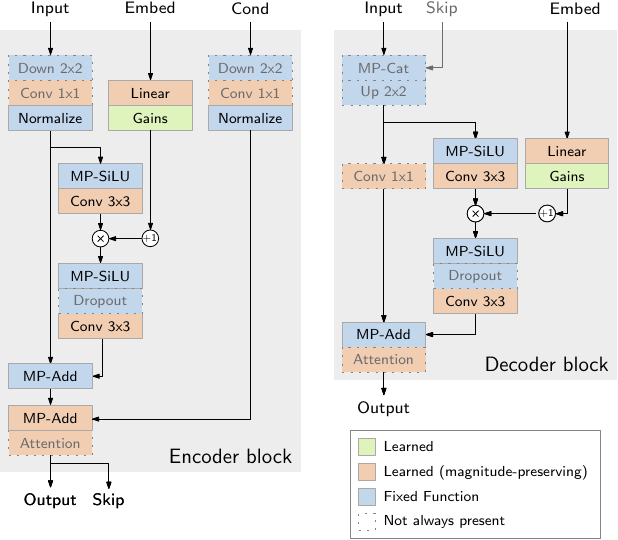}
    \caption{Encoder and decoder blocks of the magnitude-preserving ADM architecture~\cite{karras2024analyzing}. We modify the block-wise fusion of the conditioner via a learnable MP-Add operation, see \eqref{eq:mp_sum}.}
    \label{fig:mp_adm}
\end{figure}

We adopt the \ac{MP-ADM} architecture~\cite{karras2024analyzing} with magnitude-preserving learned layers (Sec.~\ref{sec:magnitude_preserving_layers}; see also Fig.~\ref{fig:mp_adm}). To inject conditioning, we add the noisy speech (or its downsampled version at the corresponding resolution) to the feature maps immediately after merging the residual branch in each block, using a magnitude-preserving fusion:
\begin{equation}
\label{eq:mp_sum}
\operatorname{MP\text{-}Add}(\mathbf{a},\mathbf{b},\tau)
= \frac{(1-\tau)\,\mathbf{a} + \tau\,\mathbf{b}}
{\sqrt{(1-\tau)^2 + \tau^2}},
\end{equation}
where \(\tau\in[0,1]\) is a learned interpolation coefficient (implemented via a sigmoid). This preserves feature magnitudes while allowing the model to learn how strongly to condition on \(\mathbf{y}\) at each block.

\section{Experimental Setup}\label{sec:experiments}

Following \cite{richter2023speech}, we use the same amplitude-compressed \ac{STFT} representation as input, given by $0.15\,|x|^{0.5} e^{i\angle(x)}$ for each coefficient $x\!\in\!\mathbb C$.
The forward and reverse SDEs in \eqref{eq:sb_forward_sde} and \eqref{eq:sb_reverse_sde} are parameterized with $\mathbf{f}\!=\!0$ and 
$g(t)\!=\!\sqrt{c}k^t$ where $c\!=\!0.4$ and $k\!=\!2.6$.  
For numerical stability, we set $t_\text{eps}\!=\!0.02$, which corresponds to the step size of our uniform discretization scheme with 50 sampling steps.  

All experiments are trained using two graphics processing units (NVIDIA RTX A6000) with a batch size of 16 until convergence.  
Model checkpoints are saved once every 1,024k training samples (equivalently, once every 64k training steps with a batch size of 16).  
We employ an inverse square root learning rate decay schedule~\cite{kingma2015adam}, starting from an initial rate of $2.5\!\times\!10^{-3}$ that decreases after processing $3\!\times\!10^4$ training samples.  
This explicit decay follows~\cite{karras2024analyzing}, which shows that in weight-normalized networks the implicit learning-rate decay caused by growing weight magnitudes is removed, making it necessary to schedule the learning rate directly over the course of training. Employing such a schedule improves stability and enables the optimizer to converge more reliably.

We train a total of eight models covering all combinations of three factors.  
First, models are trained on either the VoiceBank-DEMAND~\cite{valentini2016investigating} or the EARS-WHAM (v2)~\cite{richter2024ears}.  
Second, the skip connection coefficient is set to either $c_{\text{s}}\!=\!1$ or $c_{\text{s}}\!=\!0$.  
Third, the auxiliary loss weighting parameter is set to either $\alpha\!=\!0$ or $\alpha\!=\!0.001$.  
This full factorial design enables us to systematically evaluate the impact of each component on performance.

For preconditioning, we estimate the signal variances $\sigma_x^2$ and $\sigma_n^2$ as mean values from the amplitude-compressed training spectrograms.  
For VoiceBank-DEMAND, we obtain $\sigma_x^2\!=\!0.402$ and $\sigma_n^2\!=\!0.342$; for EARS-WHAM, $\sigma_x^2\!=\!0.368$ and $\sigma_n^2\!=\!0.353$.  
These values are fixed before training and kept constant at inference.

To assess the speech enhancement performance, we report a set of metrics commonly used in the field.  
We compute the \ac{SI-SDR}~\cite{leroux2018sdr} to quantify the separation performance in the time domain.  
Perceptual quality is evaluated using the \ac{PESQ} score~\cite{rixPerceptualEvaluationSpeech2001}. 
We include the non-intrusive speech quality assessment models NISQA~\cite{mittag2021nisqa} and DNSMOS~\cite{reddy2021dnsmos}  to capture perceptual aspects not directly reflected in reference-based metrics.  

\begin{figure}
    \centering
    \includegraphics[width=0.77\linewidth]{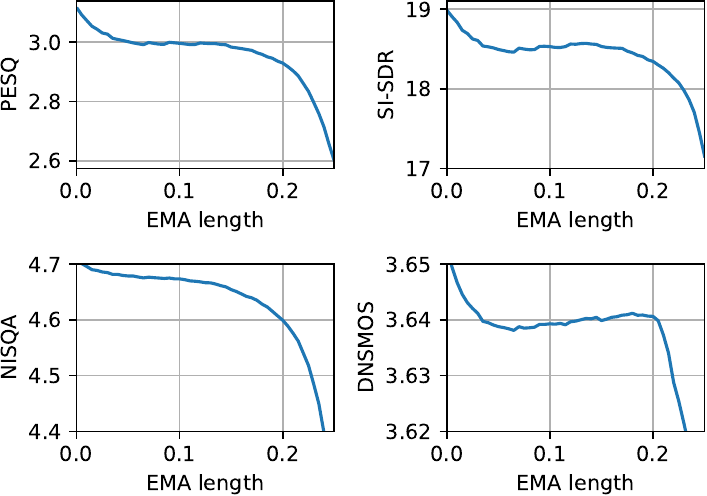}
    \caption{Speech enhancement performance on the VoiceBank-DEMAND validation set as a function of post-training EMA length.}
    \label{fig:ema_ablation}
    \vspace{-1em}
\end{figure}

\begin{table}[t]
\caption{Speech enhancement results for different skip scalings and auxiliary losses on VoiceBank-DEMAND, for matched (VoiceBank-DEMAND) and mismatched (EARS-WHAM) training.}
\begin{center}
\scalebox{0.75}{
\begin{tabular}{@{}c@{\hspace{0.7em}}lcccc@{}}
\toprule
& & SI-SDR [dB] & PESQ & DNSMOS & NISQA \\
\midrule
\multirow{4}{*}{\rotatebox{90}{\scriptsize{Matched}}} 
 & $c_\text{s}=1$, $\alpha=0.001$ & $17.50$ & $\mathbf{2.97}$ & $3.50$ & $4.70$ \\
 & $c_\text{s}=1$, $\alpha=0.0$   & $17.58$ & $2.91$ & $3.52$ & $4.71$ \\
 & $c_\text{s}=0$, $\alpha=0.001$ & $\mathbf{18.07}$ & $2.90$ & $\mathbf{3.55}$ & $\mathbf{4.76}$ \\
 & $c_\text{s}=0$, $\alpha=0.0$   & $18.04$ & $2.89$ & $\mathbf{3.55}$ & $4.75$ \\
\midrule
\multirow{4}{*}{\rotatebox{90}{\scriptsize{Mismatched}}}
 & $c_\text{s}=1$, $\alpha=0.001$ & $14.79$ & $2.69$ & $\mathbf{3.55}$ & $4.42$ \\
 & $c_\text{s}=1$, $\alpha=0.0$   & $15.71$ & $2.81$ & $3.54$ & $4.45$ \\
 & $c_\text{s}=0$, $\alpha=0.001$ & $14.23$ & $2.64$ & $3.54$ & $4.34$ \\
 & $c_\text{s}=0$, $\alpha=0.0$   & $15.18$ & $2.71$ & $\mathbf{3.55}$ & $4.48$ \\
\bottomrule
\end{tabular}
}
\end{center}
\vspace{-1.5em}
\label{tab:skip_aux_results}
\end{table}

\begin{table}[t]
\vspace{-10pt}
\caption{Speech enhancement performance on the VoiceBank-DEMAND test set. Values indicate mean and standard deviation.}
\begin{center}
\scalebox{0.75}{
\begin{tabular}{@{}c@{\hspace{0.7em}}lcccc@{}}
\toprule
& & SI-SDR [dB] & PESQ & DNSMOS & NISQA \\
\midrule
& Clean & - & $4.64 \pm 0.00$ & $3.55 \pm 0.28$ & $4.50 \pm 0.30$ \\
& Noisy & $8.44 \pm 5.61$ & $1.97 \pm 0.75$ & $3.09 \pm 0.39$ & $3.03 \pm 0.82$ \\
\midrule
\multirow{4}{*}{\rotatebox{90}{\scriptsize{Matched}\qquad}} 
& SGMSE+~\cite{richter2023speech} & $17.35 \pm 3.33$ & $2.93 \pm 0.62$ & $3.56 \pm 0.28$ & $4.51 \pm 0.38$ \\
& + (no EMA) & $17.64 \pm 3.14$ & $2.81 \pm 0.58$ & $3.53 \pm 0.28$ & $4.34 \pm 0.32$ \\
& SB-VE~\cite{jukic2024schr} & $19.41 \pm 3.48$ & $2.91 \pm 0.76$ & $3.59 \pm 0.30$ & $4.70 \pm 0.39$ \\
& + (no EMA) & $\mathbf{19.53 \pm 3.45}$ & $2.87 \pm 0.73$ & $\mathbf{3.59 \pm 0.29}$ & $4.66 \pm 0.38$ \\
& EDM2SE & $17.50 \pm 2.63$ & $\mathbf{2.97 \pm 0.71}$ & $3.50 \pm 0.31$ & $\mathbf{4.70 \pm 0.34}$ \\
\midrule
\multirow{4}{*}{\rotatebox{90}{$\quad$\scriptsize{Mism.}}} 
& SGMSE+~\cite{richter2023speech} & $10.13 \pm 5.68$ & $2.62 \pm 0.60$ & $3.51 \pm 0.29$ & $4.52 \pm 0.33$ \\
& SB-VE~\cite{jukic2024schr} & $17.71 \pm 4.05$ & $2.00 \pm 0.61$ & $3.56 \pm 0.29$ & $4.32 \pm 0.56$ \\
& EDM2SE & $14.79 \pm 3.05$ & $2.69 \pm 0.63$ & $3.55 \pm 0.31$ & $4.42 \pm 0.47$ \\
\bottomrule
\end{tabular}
\label{tab:2}
}
\end{center}
\vspace{-1.5em}
\end{table}

\begin{table}[t]
\caption{Speech enhancement performance on EARS-WHAM (v2) test set in 16$\,$kHz. Models marked with $^*$ were trained at 48\,kHz and require upsampling of test files before enhancement and downsampling after processing. Values indicate mean and standard deviation.}
\begin{center}
\scalebox{0.75}{ 
\begin{tabular}{lcccc}
\toprule
 & SI-SDR [dB] & PESQ & DNSMOS & NISQA \\
\midrule
Clean & - & $4.64 \pm 0.00$ & $3.89 \pm 0.28$ & $4.09 \pm 0.83$ \\
Noisy & $5.36 \pm 5.90$ & $1.24 \pm 0.21$ & $2.73 \pm 0.31$ & $1.95 \pm 0.71$ \\
\midrule
SGMSE+$^*$~\cite{richter2023speech} & $14.52 \pm 5.07$ & $\mathbf{2.19 \pm 0.59}$ & $\mathbf{3.79 \pm 0.29}$ & $\mathbf{4.08 \pm 0.80}$ \\
SB-VE$^*$~\cite{jukic2024schr} & $12.40 \pm 5.57$ & $1.49 \pm 0.35$ & $3.54 \pm 0.36$ & $3.37 \pm 0.83$ \\
EDM2SE & $\mathbf{14.77 \pm 3.69}$ & $2.14 \pm 0.61$ & $3.74 \pm 0.32$ & $3.94 \pm 0.86$ \\
\bottomrule
\end{tabular}
\label{tab:3}
}
\end{center}
\vspace{-1.5em}
\end{table}

\section{Results}\label{sec:results}

We first study the effect of the \ac{EMA} length using the model with $c_{\text{s}}\!=\!1$ and $\alpha\!=\!0.001$, trained on VoiceBank-DEMAND~\cite{valentini2016investigating}.
Fig.~\ref{fig:ema_ablation} shows the validation set performance versus EMA length.
Shorter \ac{EMA} lengths consistently outperform longer ones across all metrics.
Performance then plateaus before dropping sharply once $\sigma_{\text{rel}}\!>\!0.2$.
This differs from image generation, where larger EMA lengths are often preferred~\cite{karras2024analyzing}.
While EMA can improve system-level metrics such as the \ac{FID} in image synthesis, our results indicate that longer EMA lengths harm instance-level metrics in speech enhancement.
We therefore fix $\sigma_{\text{rel}}\!=\!0.001$ (the smallest possible EMA length using the post-training reconstruction method) for all subsequent experiments.

We next compare skip configurations, auxiliary loss weights, and matched vs. mismatched training-test conditions.
Table~\ref{tab:skip_aux_results} summarizes results on the VoiceBank-DEMAND test set.
Skip scaling affects the metrics differently: $c_\text{s}\!=\!1$ (noise prediction) favors PESQ, whereas $c_\text{s}\!=\!0$ (clean speech prediction) slightly improves SI-SDR and non-intrusive measures.
Using an $\ell_1$ loss in the time-domain gives small gains in matched conditions but degrades mismatched performance.
Based on the best PESQ scores in the matched condition, we adopt $c_\text{s}\!=\!1$ and $\alpha\!=\!0.001$ for the next experiments.

Table~\ref{tab:2} compares EDM2SE with SGMSE+~\cite{richter2023speech} and SB-VE~\cite{jukic2024schr}.
Overall, EDM2SE performs on par with the other models, with no statistically significant differences across metrics.
Under mismatched conditions, however, EDM2SE shows enhanced robustness, highlighting the method's strong generalization capability.
For completeness, we also evaluate SGMSE+ and SB-VE without EMA, which yields slightly higher SI-SDR scores at the cost of small drops in the other metrics.
This indicates that EMA may also not be critical in non-magnitude-preserving architectures such as NCSN++. 
However, the training dynamics may differ since the baselines do not use an explicit learning-rate schedule but rely on the Adam optimizer with momentum, which could influence convergence behavior. Investigating the interaction between EMA, learning-rate scheduling, and the Adam optimizer is beyond the scope of this paper but represents an interesting direction for future work.

Finally, Table~\ref{tab:3} reports results on the EARS-WHAM (v2) test set at 16$\,$kHz.
For SGMSE+ and SB-VE, we use pretrained checkpoints from~\cite{richter2024diffusion}, trained at 48$\,$kHz. Therefore, test files are upsampled before enhancement and downsampled afterward.
On this dataset, EDM2SE and SGMSE+ perform similarly, while SB-VE struggles--likely due to the resampling procedure.

\section{Conclusion}\label{sec:conclusion}

We presented \ac{EDM2SE}, a diffusion-bridge speech enhancement method based on the \ac{EDM2} framework using the adapted \ac{MP-ADM} architecture.
Our approach combines time-dependent input/output preconditioning with two complementary skip-connection configurations, enabling the neural network to predict either environmental noise or clean speech.
This design stabilizes training, controls activation magnitudes, and exposes a trade-off between signal-to-distortion ratio and perceptual scores.

For the first time, we examined the role of EMA in diffusion-based speech enhancement by approximating different EMA profiles post-training.
In contrast to image generation, we found that longer EMA horizons can harm instance-level metrics, whereas shorter or absent EMA may yield better performance.

Across VoiceBank-DEMAND and EARS-WHAM, our method achieves competitive speech enhancement performance and exhibits strong robustness under mismatched conditions.
These findings provide new insights into EMA behavior, magnitude preservation, and skip-connection design, and highlight promising directions for future work on generalizable diffusion-based speech enhancement.

\bibliographystyle{IEEEtran}
\bibliography{refs}

\end{document}